\documentclass[11pt,letterpaper]{article}
\usepackage{graphicx}
\usepackage{amsmath}
\usepackage{osajnl2}
\usepackage[draft]{hyperref}
\begin{document}

\title{Theory of Fiber Optic Raman Polarizers}

\author{Victor V. Kozlov$^{1,2}$, Javier Nu$\bar{\hbox{n}}$o$^3$,
Juan Diego Ania-Casta$\tilde{\hbox{n}}$\'{o}n$^3$, and Stefan Wabnitz$^{1}$}
\address{$^1$Department of Information Engineering, Universit\`{a} di
Brescia, Via Branze 38, 25123 Brescia, Italy\\
$^2$Department of Physics, St.-Petersburg State University,
Petrodvoretz, St.-Petersburg, 198504, Russia\\
$^3$Instituto de Optica, Consejo Superior de Investigaciones
Cientificas (CSIC), 28006 Madrid, Spain}

\email{victor.kozlov@email.com}

\begin{abstract}
The theoretical description of a Raman amplifier based on the vector model of
randomly birefringent fibers is proposed and applied to the characterization
of Raman polarizers. The Raman polarizer is a special type of Raman
amplifier with the property of producing a highly repolarized beam when fed by 
relatively weak and unpolarized light.
\end{abstract}


Polarization-dependent gain (PDG), an intrinsic characteristic
of optical fiber-based Raman amplifiers, is generally 
considered an unwanted feature for telecom-related applications. 
Very recently such opinion about the role of PDG was reversed, 
as the quest for higher transmission capacities brings to the 
forefront the need for polarization multiplexing protocols and 
polarization-controlling devices. Indeed, Martinelli {\it et. al.} 
demonstrated in Ref.~\cite{martinelli} such a device, called 
Raman polarizer, which selectively amplifies only one polarization 
mode of the input beam, and thereby yields only this mode at 
the output, independently of the input state of polarization (SOP) 
of the signal beam. The development of a simple, yet rigorous 
as well as computer-friendly theory of Raman polarizers along 
with the scheme for their characterization is thus the purpose 
of this Letter.

Telecom fibers are randomly birefringent fibers. Representative 
examples of vector theories of Raman amplifiers developed 
for telecom fibers can be found in Refs.~\cite{agrawal2,galtarossa}. 
The analytic theory of Ref.~\cite{agrawal2} is limited by the 
condition that the beat length $L_B$ is smaller than the 
birefringence correlation length $L_c$, and therefore their 
validity is questionable when applied to Raman polarizers, 
which as we shall see require the opposite inequality 
$L_B\gg L_c$. The full-scale numerical approach in 
Ref.~\cite{galtarossa} accurately models a randomly 
birefringent fiber as consisting of fiber spans with
randomly distributed values and orientations of the birefringence. 
Typically, thousands of such realizations are required for getting 
an accurate statistics. Hence the required computer time is three 
to four orders of magnitude longer than for the numerical modeling 
involved in the theory presented below. In addition to the much 
faster performance, our theory is formulated in terms of a set of 
deterministic differential equations, and as such allows for a simple
physical interpretation.

Starting with the equations of motion formulated by Lin and 
Agrawal in Ref.~\cite{agrawal2} we extend the one-beam model 
of the stochastic fiber proposed by Wai and Menyuk in 
Ref.~\cite{wai_menyuk} to two beams interacting not only via 
Kerr, but also via Raman effect.  Detailed derivations can be 
found in Ref.~\cite{archive}, while here we only provide the final 
equation formulated for the Stokes vector 
$S^{(s)}=(S_1^{(s)},\, S_2^{(s)},\, S_3^{(s)})$ of the signal beam:
\begin{eqnarray}
&& \left(\partial_z+\beta^\prime (\omega_s)\partial_t\right)
S^{(s)}=-\alpha_sS^{(s)}+
\nonumber\\
&& \gamma (\omega_s) \left(S^{(s)}\times \boldsymbol{J}_S^{(s)}(z) 
S^{(s)}+S^{(s)}\times \boldsymbol{J}_X(z) S^{(p)}\right)
\nonumber\\
&& +\epsilon_sg_0
\left(S_{0}^{(p)}J_{R0}S^{(s)}
+S_0^{(s)}\boldsymbol{J}_R(z)S^{(p)}\right)\, .
\label{OL1}
\end{eqnarray}
The components of the Stokes vector are written in terms 
of the two polarization components $V_{s1}$ and $V_{s2}$ 
of the slowly varying signal field in the appropriate reference 
frame, as $S_1^{s}=V_{s1}V_{s2}^*+V_{s1}^*V_{s2}$,
$S_2^{s}=i(V_{s1}^*V_{s2}-V_{s1}V_{s2}^*)$,
$S_3^{s}=\vert V_{s1}\vert^2 -\vert V_{s2}\vert^2$. Similar equations and
definitions (with labels $p$ and $s$ interchanged) hold for the pump beam.
$\gamma (\omega_s)$ is the Kerr coefficient of the fiber at frequency
$\omega_s$ of the signal beam; $g_0$ is the Raman gain coefficient;
$\beta^\prime (\omega_s)$ is the inverse group velocity of the signal beam;
$\alpha_s$ is the attenuation coefficient; $\epsilon_s=1$;
$\epsilon_p=-\omega_p/\omega_s$.  Matrices in the Eq.~(\ref{OL1})
are all diagonal with elements
$\boldsymbol{J}_R=\hbox{diag}(J_{R1},\, J_{R2},\, J_{R3})$,
$\boldsymbol{J}_X=\hbox{diag}(J_{X1},\, J_{X2},\, J_{X3})$,
$\boldsymbol{J}_S=\hbox{diag}(J_{S1},\, J_{S2},\, J_{S3})$. Here
$J_{R1}=\langle \hbox{Re}(u_{14}^2-u_{10}^2)\rangle$,
$J_{R2}=-\langle \hbox{Re}(u_{14}^2+u_{10}^2)\rangle$,
$J_{R3}=-\langle \vert u_{14}\vert^2-\vert u_{10}\vert^2\rangle$,
$J_{X1}=\frac{2}{3}\langle \hbox{Re}(u_{10}^2+u_{13}^2-u_9^2-u_{14}^2)\rangle$,
$J_{X2}=\frac{2}{3}\langle \hbox{Re}(u_{10}^2+u_{14}^2-u_9^2-u_{13}^2)\rangle$,
$J_{X3}=\frac{2}{3}\langle \vert u_{9}\vert^2+\vert u_{14}\vert^2
-\vert u_{13}\vert^2-\vert u_{10}\vert^2\rangle$,
$J_{S1}=\frac{1}{3}\langle \hbox{Re}(u_{6}^2)\rangle$,
$J_{S2}=-\frac{1}{3}\langle \hbox{Re}(u_{6}^2)\rangle$,
$J_{S3}=\frac{1}{3}\left[3\langle u_3^2\rangle -1\right]$,
and also $J_{R0}=\langle \vert u_{10}\vert^2+\vert u_{14}\vert^2\rangle$.
The three groups of coefficients
$\{ \langle u_1^2\rangle ,\, \langle  u_2^2\rangle ,\, \langle u_3^2\rangle\}$,
$\{ \langle \hbox{Re}^2(u_4)\rangle ,\, \langle\hbox{Re}^2(u_5)\rangle ,\,
\langle \hbox{Re}^2(u_6)\rangle\}$, and $\{ \langle \hbox{Im}^2(u_4)\rangle ,\,
\langle\hbox{Im}^2(u_5)\rangle ,\, \langle \hbox{Im}^2(u_6)\rangle\}$ obey
equations
\begin{eqnarray}
&& \partial_z G_1=-2L_c^{-1}(G_1-G_2)\, ,
\nonumber\\
&& \partial_z G_2=2L_c^{-1}(G_1-G_2)-4\Delta\beta (\omega_s)G_4 
\, ,
\nonumber\\
&& \partial_z G_3=4\Delta\beta (\omega_s)G_4 \, ,
\nonumber\\
&& \partial_z G_4=-L_c^{-1}G_4+2\Delta\beta 
(\omega_s)(G_2-G_3)\, ,
\nonumber
\end{eqnarray}
when we associate them with  $\{ G_1,\, G_2,\, G_3\}$ respectively. 
Initial conditions are respectively $(1,\, 0,\, 0)$, $(0,\, 1,\, 0)$, and 
$(0,\, 0,\, 1)$. In turn, the rest four groups of coefficients
$\{ \langle \hbox{Re}^2(u_7)\rangle ,\, \langle\hbox{Re}^2(u_8)\rangle ,\,
\langle \hbox{Re}^2(u_9),\, \langle\hbox{Re}^2(u_{10})\rangle\}$,
$\{ \langle \hbox{Im}^2(u_7)\rangle ,\, \langle\hbox{Im}^2(u_8)\rangle ,\,
\langle \hbox{Im}^2(u_9),\, \langle\hbox{Im}^2(u_{10})\rangle\}$,
$\{ \langle \hbox{Re}^2(u_{11})\rangle ,\, \langle\hbox{Re}^2(u_{12})\rangle ,\,
\langle \hbox{Re}^2(u_{13}),\, \langle\hbox{Re}^2(u_{14})\rangle\}$,
and $\{ \langle \hbox{Im}^2(u_{11})\rangle ,\, \langle\hbox{Im}^2(u_{12})\rangle 
,\, \langle \hbox{Im}^2(u_{13}),\, \langle\hbox{Im}^2(u_{14})\rangle\}$,
can be found from equations
\begin{eqnarray}
&& \partial_z G_1=-2L_c^{-1}(G_1-G_2)
+2\Delta_-G_5\, ,
\nonumber\\
&& \partial_z G_2=2L_c^{-1}(G_1-G_2)
-2\Delta_+G_6\, ,
\nonumber\\
&& \partial_z G_3=2\Delta_+G_6\, ,
\nonumber\\
&& \partial_z G_4=-2\Delta_-G_5\, ,
\nonumber\\
&& \partial_z G_5=\Delta_-(G_4-G_1)
-L_c^{-1}G_5\, ,
\nonumber\\
&& \partial_z G_6=\Delta_+(G_2-G_3)
-L_c^{-1}G_6\, ,
\nonumber
\end{eqnarray}
when we associate them with $\{ G_1,\, G_2,\, G_3,\, G_4\}$,
with initial conditions as $(1,\, 0,\, 0,\, 0)$, $(0,\, 0,\, 0,\, 1)$, 
$(0,\, 1,\, 0,\, 0)$, and $(0,\, 0,\, 1,\, 0)$, respectively. Here,
$\Delta_\pm\equiv\Delta\beta (\omega_p)\pm\Delta\beta (\omega_s)$, 
where $\Delta\beta (\omega_s)$ [$\Delta\beta (\omega_p)$] is 
the magnitude of the birefringence at frequency $\omega_s$ 
($\omega_p$). The power of the signal beam defined as
$S_0^{(s)}=\left({S_1^{(s)}}^2+{S_2^{(s)}}^2+{S_3^{(s)}}^2\right)^{1/2}$ 
obeys the equation
\begin{eqnarray}
&& \left(\partial_z +\beta^\prime (\omega_s)\partial_t\right)
S_0^{(s)} =-\alpha_sS_0^{(s)} +
g_0\Big(J_{R0}S_0^{(s)}S_0^{(p)}
\nonumber\\
&&
+J_{R1}S_1^{(s)}S_1^{(p)}
+J_{R2}S_2^{(s)}S_2^{(p)}
+J_{R3}S_3^{(s)}S_3^{(p)}\Big)\, .
\label{OL2}
\end{eqnarray}

Eqs.~(\ref{OL1}) and (\ref{OL2}) for the signal (and pump) fields 
are the key finding of our study. These equations are valid for 
a wide range of parameters and regimes, for undepleted as 
well as with a depleted pump. The only limitation is that the 
total length of the fiber $L$ and/or the nonlinear length
$L_{NL}=[\gamma (\omega_s) S_0^{(p)}]^{-1}$ be longer than 
the correlation length $L_c$. Eqs.~(\ref{OL1}) and (\ref{OL2}) 
can be easily solved numerically, in particular in the 
co-propagating configuration and undepleted pump regime, 
which is of interest to us here. In this case the $z$-dependent 
elements on the diagonals of the SPM, XPM and Raman 
matrices, $\boldsymbol{J}_S$, $\boldsymbol{J}_X$, and
$\boldsymbol{J}_R$, are obtained as previously discussed.

\begin{figure}[htbp]
\begin{center}
\includegraphics[scale=0.9]{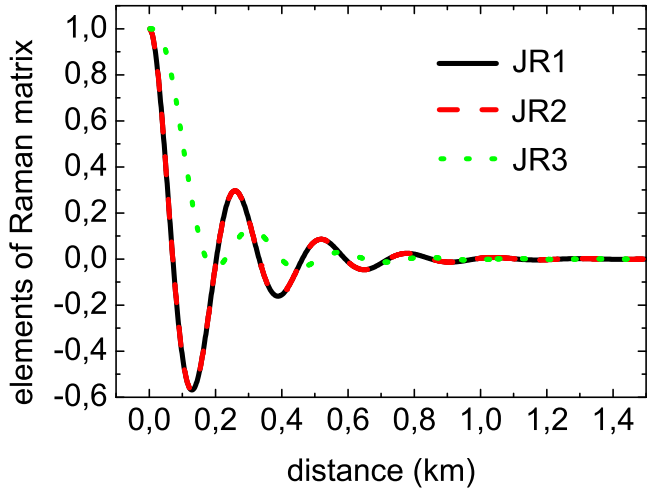}
\end{center}
\caption{Elements of the Raman matrix ($J_{R1}$ -- black solid,
$J_{R2}$ -- red dashed, and $J_{R3}$ -- green dotted) as function of
distance in the fiber for $L_B(\omega_p)=0.016$~km and 
$L_c=0.05$~km. (note that the black solid and red
dashed curves coincide, i.e. $J_{R1}=J_{R2}$.)}
\label{ris1}
\end{figure}

When doing this, we found that both SPM  and XPM effects 
have virtually no impact on the performance of Raman 
polarizers operating in the undepleted pump regime. In 
contrast, the form of the Raman matrix is of paramount 
importance. The larger the coefficients on the diagonal, 
the stronger the PDG. For moderate values of the 
polarization mode dispersion (PMD) coefficient, Raman 
diagonal terms only take appreciable values near the fiber 
input, as illustrated in Fig.~\ref{ris1}. Therefore, the power 
of the pump beam is to be high, in order to provide 
significant amplification over the first few hundreds meters 
of the fiber.

\begin{figure}[htbp]
\begin{center}
\includegraphics[scale=0.8]{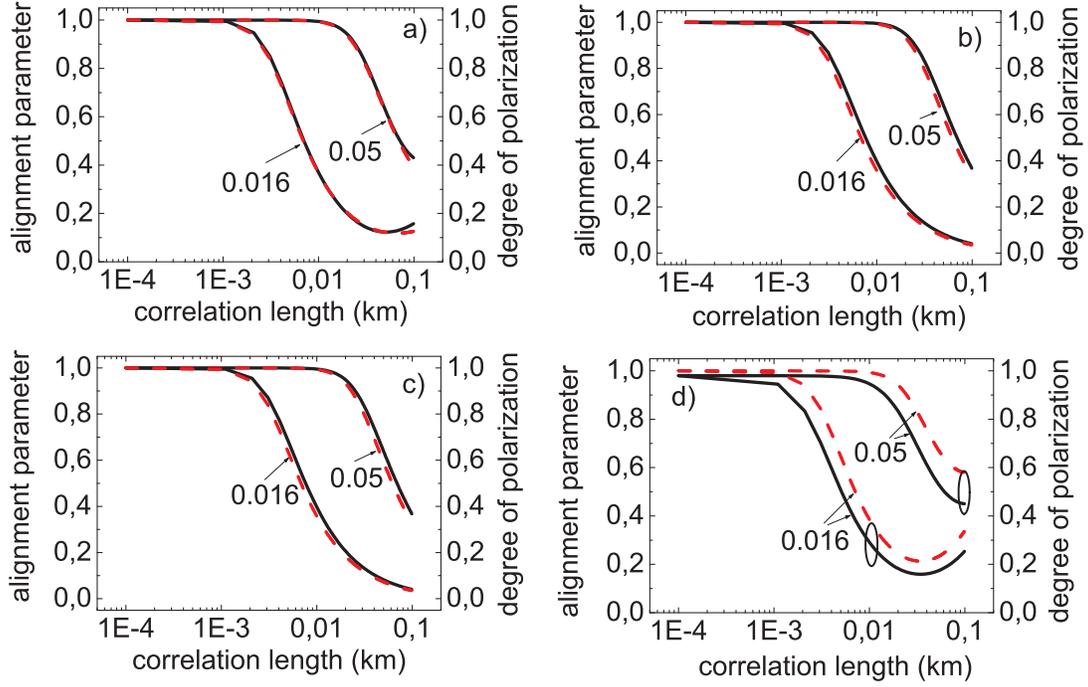}
\end{center}
\caption{DOP of the signal beam (black, solid) and alignment
parameter $A_{\uparrow\uparrow}$ (red, dashed) as function 
of correlation length $L_c$ for the four SOPs of the pump 
beam: a) $(1/\sqrt{3})(1,\, 1,\, 1)$; b) $(1,\, 0,\, 0)$; c) $(0,\, 1,\, 0)$; 
d) $(0,\, 0,\, 1)$. Here and in Figs.~\ref{ris3} and \ref{ris4}, the 
value of the beat length $L_B(\omega_p)$ is indicated on the 
plots in km. The two ellipses on plot d) indicate one (of infinitely 
many) pair of points with equal PMD coefficients. Other 
parameters are (also used in Figs.~\ref{ris3}, \ref{ris4}): input 
signal power $1$~$\mu$m; input pump power $8$~W;  
$g_0=0.6$~(W$\cdot$km)$^{-1}$; $\gamma =1$~(W$\cdot$km)$^{-1}$; 
$\alpha =0.2$~dB/km; $L=1.5$~km.}
\label{ris2}
\end{figure}

\begin{figure}[htbp]
\begin{center}
\includegraphics[scale=0.7]{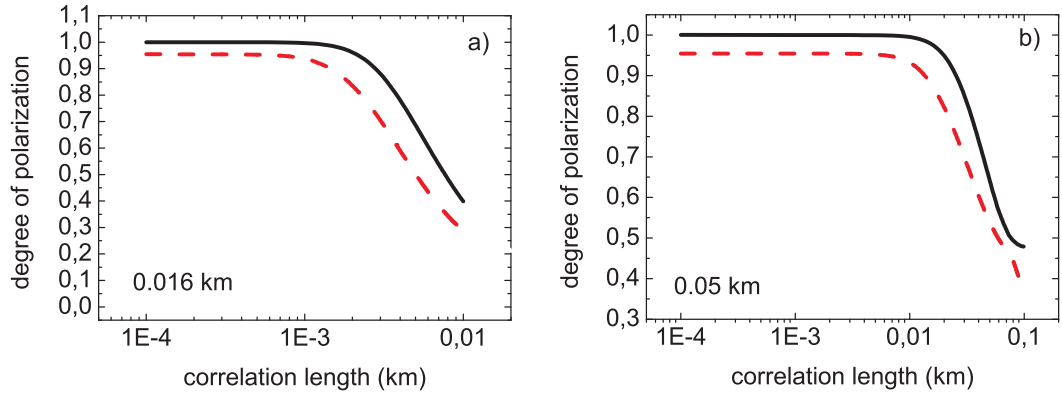}
\end{center}
\caption{DOP of the signal beam for two SOPs of the pump 
beam which either maximize (black, solid) or minimize 
(red, dashed) the signal DOP. For each value of $L_c$ 
we perform a separate search for these two SOPs.}
\label{ris3}
\end{figure}

For analyzing the performance of Raman polarizers we identify
three characteristic quantities: the degree of polarization (DOP) of the 
outcoming signal beam, its SOP, and the overall signal gain. The 
DOP and SOP characteristics are illustrated in Fig.~\ref{ris2}. Since the 
signal SOP depends on the pump SOP, it is reasonable to define a 
quantity that measures the relative difference between these two SOPs. 
As usual, such quantity is the alignment parameter
\begin{equation}
A_{\uparrow\uparrow}
\equiv\frac{\left\langle S_1^{(s)}S_1^{(p)}+S_2^{(s)}S_2^{(p)}
+S_3^{(s)}S_3^{(p)}\right\rangle}{S_0^{(s)}S_0^{(p)}}\, ,
\label{OL3}
\end{equation}
which is the cosine of the angle between the pump and the signal 
Stokes vectors, averaged over the ensemble of beams with random 
SOPs which models the unpolarized signal beam. The hypothesis 
that the signal SOP is attracted to the pump SOP is rooted in the 
model of isotropic fibers, in which $J_{R1}=J_{R2}=J_{R2}=1$.
In randomly birefringent fibers, the equality and even positivity of 
the three elements is not always the case, as exemplified in the plot of
Fig.~\ref{ris1}. In these cases, it is remarkable that the signal SOP is 
attracted to an SOP which is different from that of the pump. In 
spite of this observation, we found that for ideal Raman polarizers 
(those with DOP$>0.9$), and in the range of lengths $0.001<L_B<0.05$ 
and $0.0001<L_c<0.05$, given here in km, the signal SOP on average 
is attracted to the pump SOP, see Fig.~\ref{ris2}. This is not the case 
in the counter-propagating configuration, for which the appropriate 
alignment parameter $A_{\uparrow\downarrow}$ is different from that
given in Eq.~(\ref{OL3}), see \cite{archive}. Moreover, the performance
of Raman polarizers (namely, DOP) sensitively depends on the pump 
SOP, as demonstrated in Fig.~\ref{ris3}.

\begin{figure}[htbp]
\begin{center}
\includegraphics[scale=0.7]{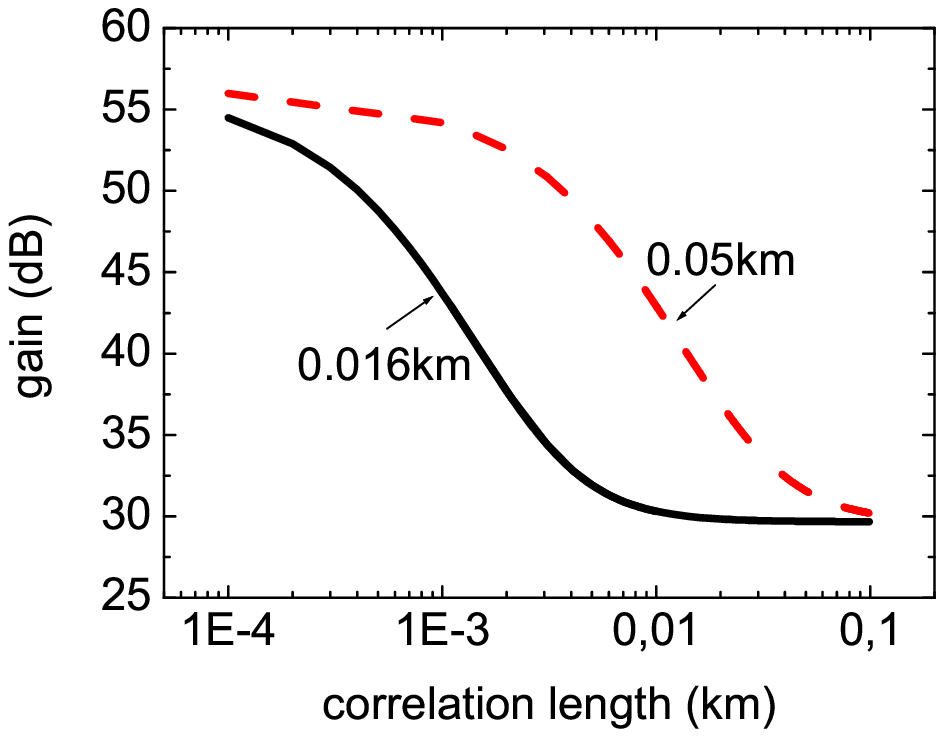}
\end{center}
\caption{Average Raman polarizer gain as a function of the correlation
length. The pump SOP is $(1,\, 0,\, 0)$ and the signal beam is initially 
unpolarized.}
\label{ris4}
\end{figure}

Another important practical issue is the selection of fibers for 
Raman polarizers. The main parameter in this selection is 
the value of the PMD coefficient. In this respect we found that for 
obtaining a signal DOP close to unity (i.e., $>0.99$) the PMD 
coefficient should be less than $0.0145$~ps$/\sqrt{\hbox{km}}$ 
for, say, $8$~W of pump power (as in Ref.~\cite{martinelli}).
Nevertheless we found that the PMD coefficient does not always 
provide full information about the fiber. For example in 
Fig.~\ref{ris2}(d) we can see that two fibers with equal PMD 
coefficients exhibit a different performance as Raman polarizers. 
In one case, the DOP is  $0.25$, in the other -- $0.45$. For this 
reason, it is preferable to consider the beat and correlation 
lengths separately, rather than combining them into the single 
PMD coefficient, which for our model is expressed as 
\cite{wai_menyuk}: $D_p=2\sqrt{2}\pi\sqrt{L_c}/(L_B\omega_s)$.

The third characteristic of Raman polarizers is Raman gain, see Fig.~\ref{ris4}.
Even for a $1.5$~km long fiber with $8$~W of pump power we may have an
enormous $55$~dB gain that is almost twice the gain of the same Raman amplifier,
but with a high value of the PMD coefficient. This means that Raman polarizers 
are simultaneously very efficient Raman amplifiers. Such values of gain are 
obtained in the undepleted regime, i.e. for input signal powers in the $\mu$W 
range. For the mW range which is typical of telecom applications the analysis 
necessarily enters the depleted pump regime, to which our theory can also be 
readily applied.

In conclusion, we presented a theory for describing the interaction of two 
optical beams in randomly birefringent fibers via Kerr and Raman effects, 
and applied it to the quantification of the performance of Raman polarizers.

We thank L. Palmieri for valuable comments.
This work was carried out in the framework of the "Scientific Research Project of
Relevant National Interest" (PRIN 2008) entitled "Nonlinear cross-polarization
interactions in photonic devices and systems" (POLARIZON), and in the framework
of the 2009 Italy-Spain integrated action "Nonlinear Optical Systems
and Devices" (HI2008-0075).


\end{document}